\documentclass{article}

\input pz.sty
\usepackage{longtable}
\usepackage{graphicx}
\usepackage[english]{babel}

\begin{document}

\title{\bf{Photometric and Spectroscopic Study of the Supergiant with an Infrared Excess V1027~Cygni}}

\author{V.P. Arkhipova, O.G. Taranova, N.P. Ikonnikova\footnote{E-mail: ikonnikova@gmail.com}, V.F. Esipov,\\
{G.V. Komissarova, V.I. Shenavrin, and M.A. Burlak}}

\date{\it{Sternberg Astronomical Institute,\\
Lomonosov Moscow State University, \\ Universitetskii pr. 13,
Moscow, 119992 Russia}}

\renewcommand{\abstractname}{ }

\maketitle

\begin{abstract}

We present the results of our $UBV$ and $JHKLM$-photometry for the
semiregular pulsating variable V1027~Cyg, a supergiant with an
infrared excess, over the period from 1997 to 2015 ($UBV$) and in
2009--2015 ($JHKLM$). Together with the new data, we analyze the
photometric observations of V1027 Cyg that we have obtained and
published previously. Our search for a periodicity in the $UBV$
brightness variations has led to several periods from $P=212^{d}$
to $P=320^{d}$ in different time intervals. We have found the
period $P=237^{d}$ based on our infrared photometry. The
variability amplitude, the light-curve shape, and the magnitude of
V1027~Cyg at maximum light change noticeably from cycle to cycle.
The deepest minimum was observed in 2011, when the amplitudes of
brightness variations in the star reached the following values:
$\Delta U$=1.$^{m}$28, $\Delta B$=1.$^{m}$10, $\Delta
V$=1.$^{m}$05, $\Delta J$=0.$^{m}$30, $\Delta H$=0.$^{m}$35,
$\Delta K$=0.$^{m}$32, $\Delta L$=0.$^{m}$26, $\Delta
M$=0.$^{m}$10. An ambiguous correlation of the $B-V$ and $U-B$
colors with the brightness has been revealed. For example, a
noticeable bluing of the star was observed during the deep 1992,
2008, and 2011 minima, while the variations with smaller
amplitudes show an increase in $B-V$ at the photometric minima.
The spectral energy distribution for V1027~Cyg from our photometry
in the range 0.36 ($U$)-5.0 ($M$) $\mu$m corresponds to spectral
types from G8I to K3I at different phases of the pulsation cycle.
Low-resolution spectra of V1027 Cyg in the range
$\lambda$4400--9200 \AA\ were taken during 16 nights over the
period 1995--2015. At the 1995 and 2011 photometric minima the
star's spectrum exhibited molecular TiO bands whose intensity
corresponded to spectral types M0--M1, while the photometric data
point to a considerably earlier spectral type. We hypothesize that
the TiO bands are formed in the upper layers of the extended
stellar atmosphere. We have measured the equivalent widths of the
strongest absorption lines, in particular, the infrared Ca~II
triplet in the spectrum of V1027~Cyg. The calcium triplet (Ca T)
with $W_{\lambda}(\mathrm{Ca~T})=20.3\pm1.8$ \AA\ as a luminosity
indicator for supergiants places V1027 Cyg in the region of the
brightest G--K supergiants. V1027 Cyg has been identified with the
infrared source IRAS~20004+2955 and is currently believed to be a
candidate for post-AGB stars. The evolutionary status of the star
and its difference from other post-AGB objects are discussed.

 {\it {Keywords}}: post-AGB stars, photometric observations, photometric
variability, spectroscopic observations, spectral classification,
evolution.

 \end{abstract}

 \newpage

 \section*{INTRODUCTION}

The bright ($V \approx8.^{m}65$; Hog et al. 2000) variable star
V1027 Cyg (BD+29$^{\circ}$3865=HD~333385) of spectral type K7
(according to the HD catalogue) had remained a poorly studied
object until the mid-1980s. After the detection of a considerable
far-infrared (IR) (12--100 $\mu$m) excess in the star with the
IRAS space telescope, the interest in it has increased
significantly. Volk and Kwok (1989) identified V1027 Cyg with
IRAS~20004+2955, attributed its IR excess to the radiation from
the dust shell formed during a large-scale mass loss on the
asymptotic giant branch (AGB), and classified the star as a
candidate for post-AGB objects, a star at a late evolutionary
stage on its way from AGB stars to planetary nebulae. Subsequent
studies, in particular, the analysis of the chemical composition
of the stellar atmosphere performed by Klochkova et al. (2000),
confirmed that V1027 Cyg belongs to the class of post-AGB stars.
Judging by the chemical composition of the dust shell, the star
belongs to oxygen-rich objects (Hrivnak et al. 1989; Vandenbussche
et al. 2002; He et al. 2014).

Post-AGB objects manifest themselves as supergiants with an IR
excess in a wide range of spectral types, from late K to early B.
Photometric variability that can be due to various factors has
been detected in many of them. In particular, semiregular
brightness variations caused by pulsations are typical of yellow
supergiants. In addition, the photometric behavior of post-AGB
supergiants can be affected by a variable stellar wind as well as
by changes in the parameters of the star and the circumstellar
shell related to rapid post-AGB evolution. All these facts make a
photometric monitoring of stars of this class on a time scale of
days, months, years, and decades very topical.

The photographic brightness variability in the star
BD+29$^{\circ}$3865 from 10.$^{m}$5 to 11.$^{m}$5 in 1948--1959
was detected by Wachmann (1961). Based on these observations, the
star was included in the General Catalog of Variable Stars, where
it was designated V1027 Cyg and was attributed to the class of
slow irregular L-type variables (Kukarkin et al. 1971). Our $UBV$
observations in 1991--1992 (Arkhipova et al. 1991, 1992) and,
subsequently, in 1992--1996 (Arkhipova et al. 1997) revealed
quasi-periodic brightness variations in V1027~Cyg with maximum
amplitudes up to 1.$^{m}$0 in $UBV$ and a cycle duration of
200-250$^{d}$. Radial velocity measurements for V1027~Cyg showed
that there is a phase shift between the radial velocity curve and
the light curves, with the star's maximum contraction occurring on
the ascending branch of its deep minimum (Arkhipova et al. 1997).

The IR photometry for V1027 Cyg over 1991--2008 showed
quasi-periodic $JHKLM$ brightness variations in the star with
amplitudes no greater than 0.$^{m}$2 in all bands and a period
$P=237^{d}$. The mean near-IR brightness of the star over 18 years
experienced no statistically significant changes (Taranova et al.
2009; Bogdanov and Taranova 2009). Based on their IR photometry
supplemented with the mid- and far-IR fluxes, Bogdanov and
Taranova (2009) computed the model of a spherically symmetric dust
shell composed of silicate particles.

The optical spectroscopy for V1027~Cyg has been performed
repeatedly. The spectral type of the star was estimated to be K0Ia
(Roman 1973), G7Ia (Keenan and McNeil 1976), $\sim$G7Iàb (Hrivnak
et al. 1989), and K2--4I (Winfrey et al. 1994). High-resolution
spectroscopy from Klochkova et al. (2000) allowed the fundamental
parameters ($T_{eff}$=5000 K, $\log g$=1.0), metallicity
([Fe/H]=--0.2 dex), and atmospheric chemical composition of V1027
Cyg to be determined.

The goal of this paper is to analyze the photometric and
spectroscopic variability of V1027 Cyg in the optical and IR
ranges based on our own long-term observations and to study the
properties of the star as a post-AGB object.

\section*{$UBV$ PHOTOMETRY FOR V1027 Cyg}

We began the $UBV$ observations of V1027~Cyg in 1991. The results
of its observations over 1991--1996 were published in Arkhipova et
al. (1991, 1992, 1997). After 1996 we continued the observations
of V1027 Cyg with a 60-cm Zeiss reflector at the Crimean Station
of the Sternberg Astronomical Institute with a photon-counting
photoelectric photometer (Lyutyj 1971). The observations were
carried out with a 14$''$ aperture to completely eliminate the
influence of the neighboring faint blue star located at a distance
of $\sim$25$''$ from V1027 Cyg. BD+29$^{\circ}$3871 (Sp = K4.5V;
Orosz et al. 2001), whose magnitudes $U$=11.$^{m}$44,
$B$=9.$^{m}$65, $V$=8.$^{m}$20  were determined by tying to
photometric standards with an accuracy of 0.$^{m}$01-0.$^{m}$02,
served as a comparison star. Comparison of BD+29$^{\circ}$3871
with check stars revealed no variations in its brightness
exceeding the measurement errors. The $\it{Hipparcos}$ photometric
observations in 1989--1993 did not reveal any photometric
variability of the comparison star either.

From 1997 to 2015 we obtained a total of 491 magnitude estimates
for V1027 Cyg in $UBV$. The observational errors were, on average,
$0.^{m}$01 in $B$ and $V$ and $0.^{m}$02 in $U$.

Figure 1 presents the light curves of V1027 Cyg from 1991 to 2015,
where the observations that we obtained and published previously,
along with the new data, are shown.


\begin{figure}

\includegraphics[scale=1.5]{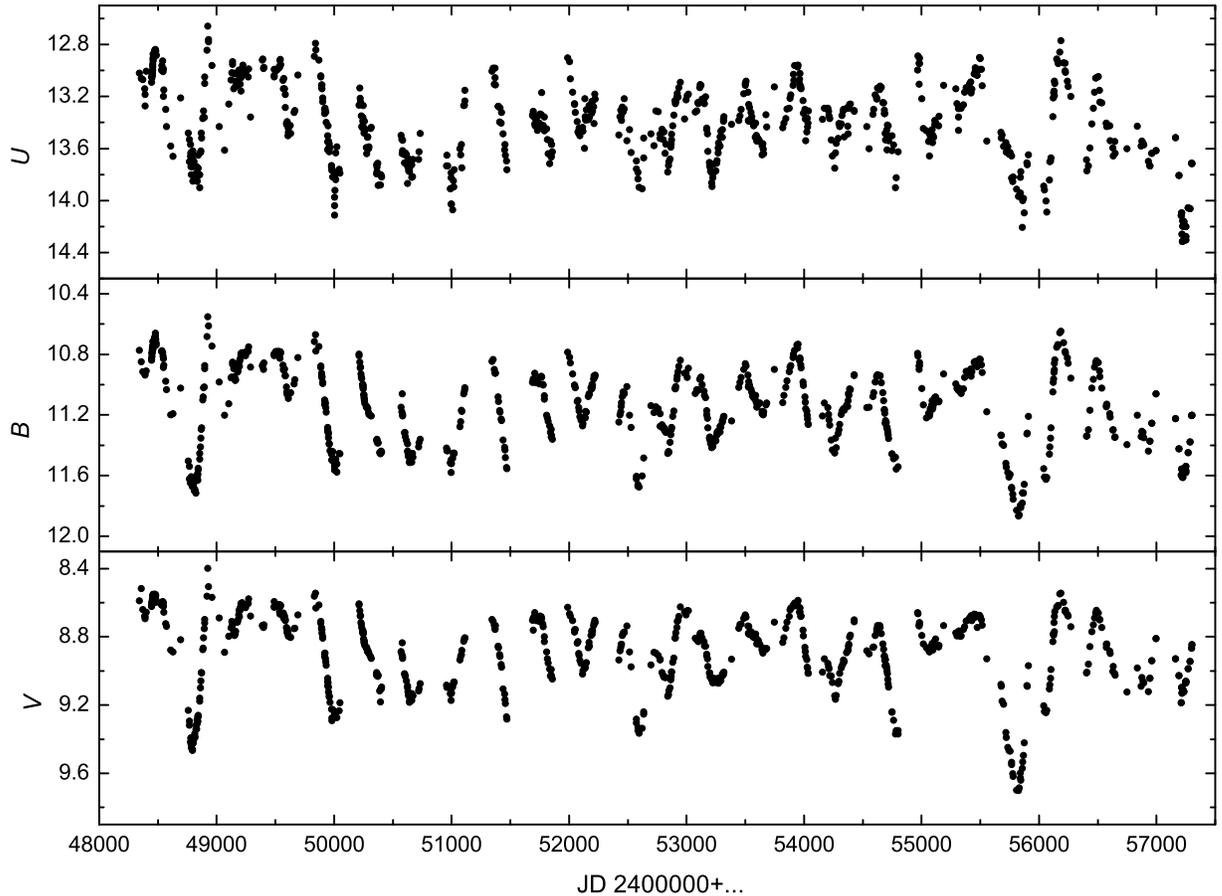}
\caption{$UBV$ light curves of V1027 Cyg from our Crimean
observations from 1991 to 2015.} \label{fig1}
\end{figure}


In 1991--2015 the star exhibited semiregular brightness
variations. The variability amplitude, the light-curve shape, and
the magnitude of V1027 Cyg at maximum light changed noticeably
from cycle to cycle. We observed both small-amplitude variations
with $\Delta V$=0.$^{m}$2 in 1994, 2005, and 2009 and variations
with depths up to 0.$^{m}$6 as well as deep photometric minima
with an amplitude up to 1$^{m}$ in 1992 and 2011. The star has
maximum variability amplitudes in $U$. The brightness level at
maxima varied within the ranges: $V_{max}=8.^{m}6\pm 0.^{m}1$,
$B_{max}=10.^{m}85\pm 0.^{m}25$, $U_{max}=13.^{m}0\pm 0.^{m}3$.
The time scale of brightness variations was 200-300$^{d}$ and was
comparable to the duration of observing seasons ($\Delta
T=230\pm55^{d}$).

\subsection*{Relationship between the Brightness and Colors}

Figure 2 shows the $V$ light and $B-V$ and $U-B$ color curves over
the entire period of observations of V1027~Cyg from 1991 to 2015.
As can be seen, the colors changed significantly with brightness.
The amplitudes of the $B-V$ and $U-B$ color variations reach
0.$^{m}3$ and 0.$^{m}$45, respectively. The color, along with the
brightness, experiences quasi-periodic variations, but there is no
unambiguous correlation between them.


\begin{figure}
\includegraphics[scale=1.5]{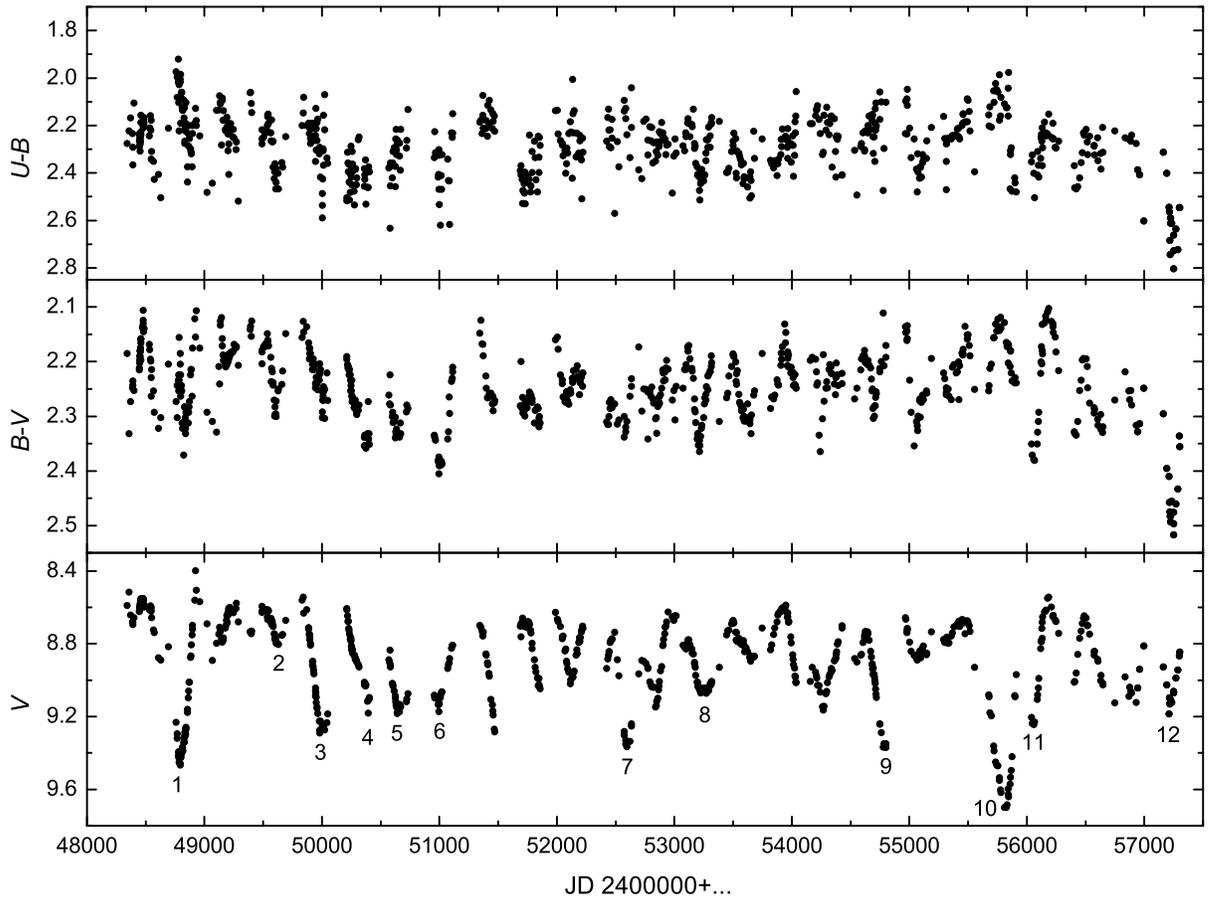}
\caption{$V$ light and $B-V$ and $U-B$ color curves from our
Crimean observations from 1991 to 2015. The numbers in the figure
mark the times of observations that are discussed in the text.}
\label{fig2}
\end{figure}


The following peculiarities of the brightness and color
variability of V1027~Cyg are noteworthy.

(1) Brightness minima close in depth can have significantly
different $B-V$ colors. Over the period 1995--1998 there were four
minima of approximately the same depth when the star faded to
$V\sim9.^{m}25$ (3, 4, 5, and 6 in Fig. 2). The $B-V$ color during
these variations as the brightness declined increased to values
that differed by $\sim 0.^{m}15$ at the minima.

(2) Close values of $B-V$ are reached at brightness minima
different in depth. The 1994 and 1995 variations (2 and 3 in Fig.
2), when at different brightness levels at the minima,
$V_{min}(1994)=8.^{m}$78 and $V_{min}(1995)=9.^{m}$26, the color
reached $B-V\sim2.^{m}3$, can serve as an example.

(3) In general, the $UBV$ light curves of V1027~Cyg are similar,
but in 2004 (8 in Fig. 2) the star exhibited variations in which
the shapes of the light curves differed in all three bands. After
the minimum was reached at JD$\sim$24~53223, the $V$ brightness
remained at $V=9.^{m}05\pm0.^{m}02$ for $\sim$ 85 days, while in
$B$ and $U$ the star lingered at its minimum for no more than 10
days. The brightness then began to rise, and the colors became
bluer as long as the star remained at its minimum in $V$: $B-V$ by
$0.^{m}15$ and $U-B$ by $0.^{m}20$.

(4) In the variations with an amplitude up to $\Delta
V\sim0.^{m}75$ the $B-V$ color generally increased as the
brightness declined. The behavior of the $U-B$ color in these
variations was ambiguous. For example, during the 1995--2002
cycles $U-B$ with slight fluctuations neither reddened nor became
bluer. A reddening of $U-B$ on the descending branches of the
light curves was observed at the 1993, 1994, 2003, 2007--2010, and
2012 minima.

Particular attention should be given to an appreciable reddening
of the star at the shallow 2015 minimum (12 in Fig. 2) when the
colors reached their maximum values over the entire period of our
observations, $B-V\approx2.^{m}5$ and $U-B\approx2.^{m}8$.

(5) The star experiences a bluing at deep brightness minima.

The deep 1992, 2008, and 2011 minima (1, 9, and 10 in Fig. 2)
showed that in its bright state the star slightly reddens as it
fades and then, starting from some brightness level, the $B-V$
color decreases and at minimum reaches the values corresponding to
the star's brightness at maximum. At the fairly deep 2002 minimum
(7 in Fig. 2) $B-V$ barely changed as the brightness declined.

The color-magnitude diagrams (Figs. 3 and 4) show different
behaviors of the colors as the brightness declines: a bluing in
some variations and a reddening in others. The fragment of the $V$
light and $B-V$ and $U-B$ color curves for 2011--2012 (Fig. 5),
where the "blue"\ and "red"\ photometric minima are adjacent,
convincingly illustrates that the star is characterized by at
least two types of variations for which the color variations are
different.

The pulsations of V1027 Cyg are mainly responsible for its
photometric variability. However, the complex photometric behavior
of the star cannot be explained by its pulsations alone. Below we
will investigate the nature of the star's variability by invoking
our spectroscopic data and IR observations.


\begin{figure}
\includegraphics[scale=1.0]{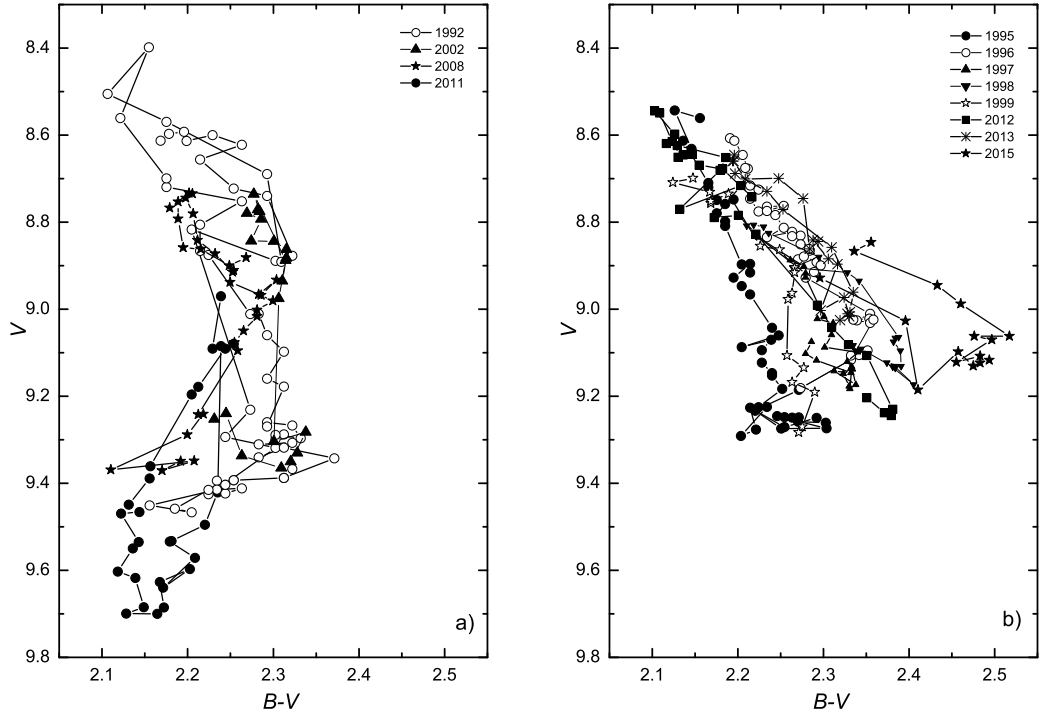}
\caption{$V-(B-V)$ diagrams for the deep (a) and shallow (b)
photometric minima of V1027 Cyg. Different symbols mark the data
for different years of observations.} \label{3}
\end{figure}



\begin{figure}
\includegraphics[scale=1.0]{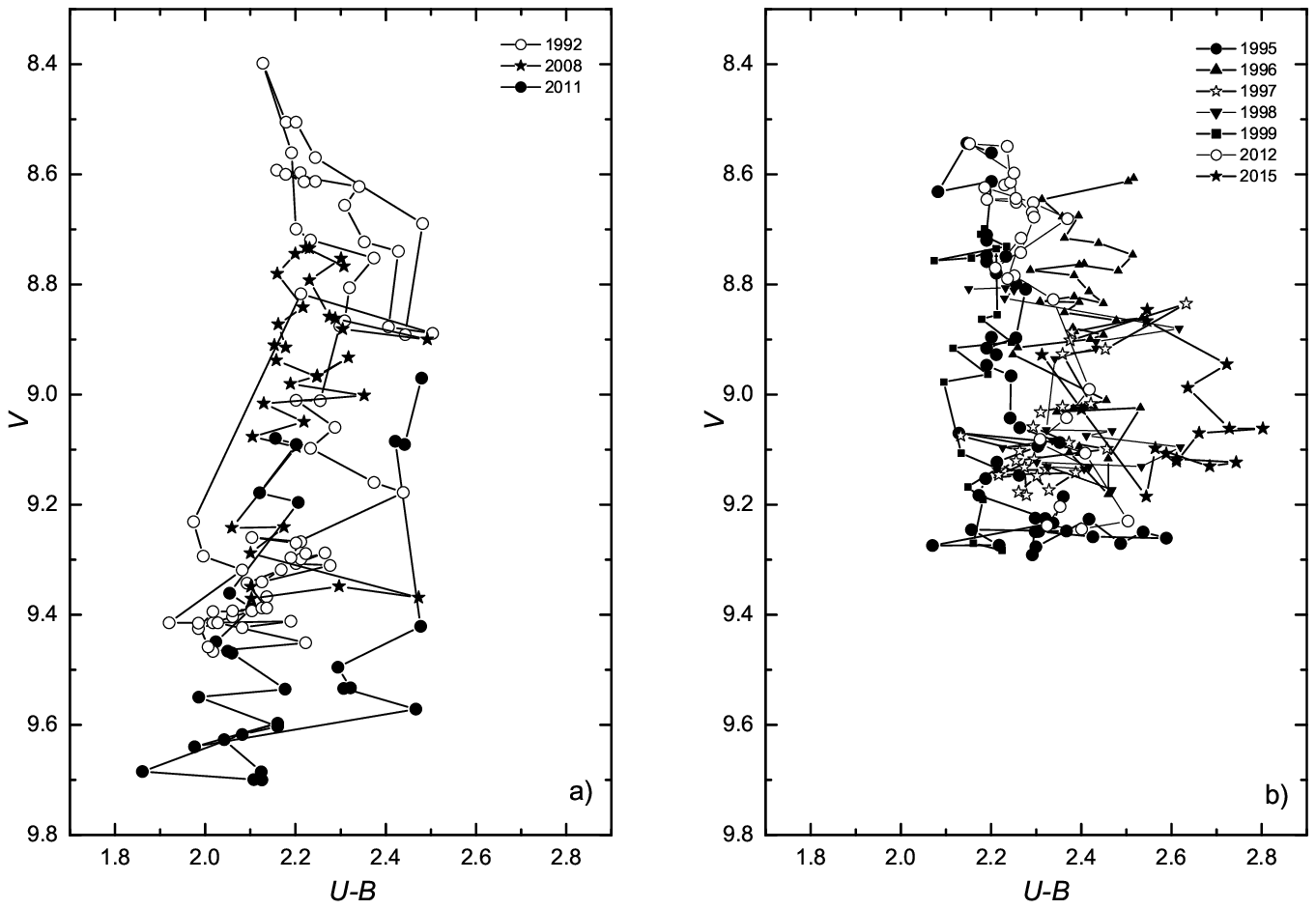}
\caption{$V-(U-B)$ diagrams for the deep (a) and shallow (b)
photometric minima of V1027 Cyg. Different symbols mark the data
for different years of observations.} \label{fig4}
\end{figure}



\begin{figure}
\begin{center}
\includegraphics[scale=1.5]{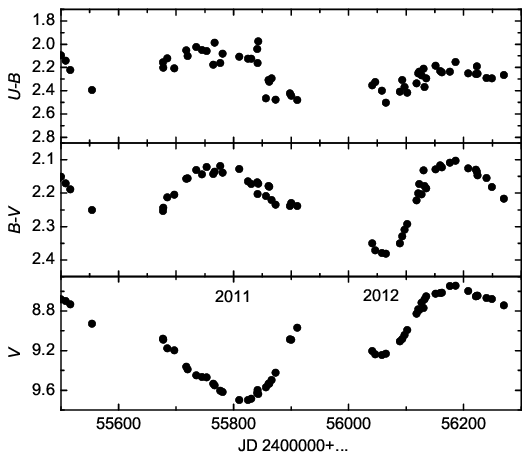}
\caption{Light and color curves in 2011--2012.} \label{fig5}
\end{center}
\end{figure}


\subsection*{Searching for a Periodicity}

Previously, we performed a Fourier analysis of our photometric
observations for V1027 Cyg in 1991--1996 and detected no prominent
peaks on the periodograms. After excluding the data referring to
the deep minima from consideration, we found a period of 256 days
(Arkhipova et al. 1997).

Here, we present a frequency analysis of our photometric $UBV$
observations for V1027~Cyg over the entire period from 1991 to
2015. The PERIOD04 code (Lenz and Breger 2005) and V.P. Goransky's
EFFECT code, which implements the Fourier transform for time
series with an arbitrary distribution of observations in time
(Deeming 1975), were used to search for a period.

Figure 6 shows the amplitude spectra of the complete series of
observations for V1027 Cyg from 1991 to 2015 in $U$ and $V$ in the
range of periods 100–-1000 days. Groups of peaks corresponding to
periods within 201--313$^{d}$ and 450–-1000$^{d}$ are identified
in the amplitude spectrum. A visual analysis of the light curves
shows that the variability time scale for V1027~Cyg does not
exceed 320$^{d}$. Therefore, the peaks in the range of periods
450--1000$^{d}$ are very likely to be a superposition of shorter
periods with one year.

Thus, the frequency analysis of our optical photometric
observations for V1027~Cyg based on the entire series of
observations revealed no variations with a dominant frequency.
Therefore, we analyzed the light curves for periodicity by
dividing the entire time of observations into intervals including
several observing seasons. Table 1 gives the time intervals, the
number of nights ($N$), the periods and amplitudes in $U$, $B$,
and $V$ corresponding to the identified frequencies in the
specified JD range. We excluded the 2011 event from consideration,
which stands out not only by the largest variability amplitude but
also by its duration: the time from maximum to minimum was more
than 300 days.


\begin{table}

\caption{Results of our search for a periodicity in the $UBV$
observations of V1027~Cyg in different time intervals}

\begin{tabular}{|c|c|c|c|c|c|c|c|}

\hline

JD&N&\multicolumn{2}{|c}{$V$}&\multicolumn{2}{|c}{$B$}&\multicolumn{2}{|c|}{$U$}
\\

\cline{3-8}
2400000+...&&$P$, d&$A$, mag&$P$, d&$A$, mag&$P$, d&$A$, mag\\
\hline

48916-49692&69&277&0.10&277&0.13&282&0.20\\
49031-51112&127&320&0.22&320&0.29&316&0.32\\
51687-55554&328&244&0.11&244&0.11&212&0.10\\
56041-57301&75&293&0.24&293&0.36&303&0.49\\

\hline

\end{tabular}

\end{table}



\begin{figure}
\includegraphics[scale=1.0]{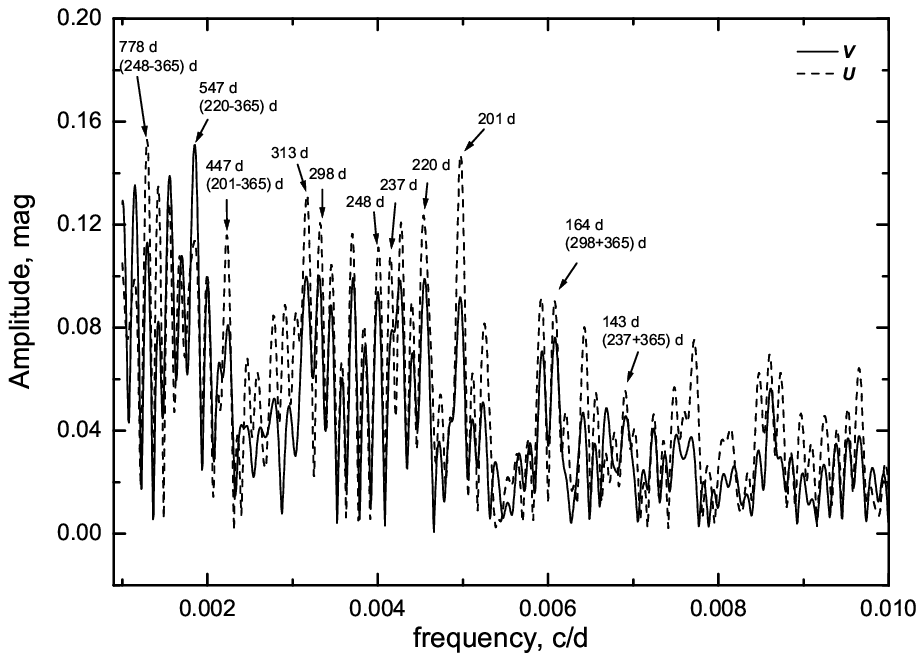}
\caption{Amplitude spectra of the complete series of observations
for V1027 Cyg from 1992 to 2015 in $U$ and $V$.} \label{fig6}
\end{figure}


Thus, our $UBV$ observations of V1027 Cyg from 1991 to 2015
revealed quasi-periodic brightness and color variations on a
variability time scale from 212 to 320 days. As will be shown
below, V1027~Cyg experiences small-amplitude variations in the
near infrared that are satisfactorily represented by a sine wave
with a period of 237 days, except for the 2011 event.

\subsection*{Extinction for V1027 Cyg}

The star lies at a low Galactic latitude ($b=-0.^{\circ}$4), and
its emission is severely distorted by interstellar extinction. The
maximum extinction toward V1027 Cyg is great, being
$E(B-V)=2.^{m}52\pm0.^{m}08$ from the maps of Schlegel et al.
(1998). For V1027 Cyg this quantity was undoubtedly overestimated.

The first extinction estimate based on $UBV$ photometry for
V1027~Cyg is given in Hrivnak et al. (1989),
$A_{V}=2.^{m}5-3.^{m}0$. The color excesses were also estimated by
Vickers et al. (2015) when modeling the spectral energy
distribution for post-AGB stars, $E(B-V)=0.^{m}73$, and by Rayner
et al. (2009), $E(B-V)=1.^{m}08$, as the difference of the
observed and normal colors for a G7 supergiant.

To estimate the extinction, we used our optical photometry
together with the data on the star's spectral type.

The published estimates of the spectral type for V1027 Cyg lie
within the range from G7I to K4I. The discrepancies may be due to
a change in the star's spectrum during its pulsations. The optical
colors also change significantly; therefore, spectroscopic and
photometric observations close in time should be used, if
possible, to properly estimate the extinction.

To estimate the color excess $E(B-V)$, we used the following:

(1) the data from Klochkova et al. (2000), where the parameters of
V1027~Cyg $T_{eff}=5000$~K è $\log g$=1.0, which correspond to the
spectral type G8I in the calibration of Strai\v{z}ys (1982), were
obtained from the spectroscopic observations on August 13, 1996
(JD=24~50309.4);

(2) the $B$ and $V$ magnitudes for JD=24~50366,
$B=11.^{m}36\pm0.^{m}01$, $V=9.^{m}02\pm0.^{m}01$ and the color
$B-V=2.^{m}34\pm0.^{m}02$;

(3) the normal color $B-V=1.^{m}27$ for a G8I star from
Strai\v{z}ys (1992).

With these input data the color excess for V1027 Cyg
$E(B-V)=1.^{m}07\pm0.^{m}02$.

It should be kept in mind that our estimate of $E(B-V)$
characterizes the total extinction, which includes both
interstellar and circumstellar components.

\section*{IR PHOTOMETRY FOR V1027~Cyg}

The IR photometry for V1027 Cyg was performed in 1991--2015 with a
1.25-m telescope at the Crimean Station of the Sternberg
Astronomical Institute using a photometer with a photovoltaic
liquid-nitrogen-cooled InSb detector. The photometer was mounted
at the Cassegrain focus of the telescope. The diameter of the
entrance aperture was $\approx 12^{\prime\prime}$, while the
spatial separation of the beams during chopping was $\approx
30^{\prime\prime}$ in the east--west direction. The star BS 7615
from the catalog by Johnson et al. (1966) served as a photometric
standard. The IR photometry for V1027 Cyg in 1991--2008 and the
computation of a spherically symmetric dust shell model were
presented in Taranova et al. (2009) and Bogdanov and Taranova
(2009). Bogdanov and Taranova (2009) highlight the following
characteristics as the main features of the IR brightness and
color variations in V1027 Cyg in 1991–-2008: first, the brightness
variations in all bands did not exceed $0.^{m}2$; second, the mean
brightness in all bands and the mean colors over this period
changed only slightly, by no more than a few hundredths of a
magnitude.

The results of our 2009–-2015 IR photometry are summarized in
Table 2. The $JKM$ brightness and $J-H$, $K-L$, and $L-M$ color
variations over the entire period of our observations (1991–-2015)
are shown in Fig. 7, where the filled and open circles refer to
the observations before and after 2009, respectively. The vertical
bars in Fig. 7 indicate the errors in the $M$ magnitude and $L-M$
color estimates; in the remaining bands the errors did not exceed
$0.^{m}01$. The solid line marks the possible linear IR brightness
and color trends.

On the whole, the pattern of IR variability in V1027 Cyg did not
change significantly over 25 years (1991--2015). However, an
unusual brightness decline whose amplitude was approximately the
same $0.^{m}25-0.^{m}30$ in $JHKL$ and $\sim0.^{m}1$ in $M$, was
observed in the IR emission from the star in 2011 near
JD~24~(55732-55810). As a result, the star's color in the range
1.25--3.5 $\mu$m did not change within 1$\sigma$; in the range
3.5--5 $\mu$m the color excess increased.

For three intervals of observations of V1027 Cyg (1991--2008,
1991--2015, and 2009--2015 (outside and at the 2011 IR brightness
minimum)) Table~3 provides the averaged IR magnitudes and colors
$\bar{m}$, their standard deviations $\sigma_{\bar{m}}$, and the
number of points N over which the data were averaged. All
quantities were corrected for extinction with $E(B-V)=1.07$. It
can be seen from Table~3 that the averaged quantities over
1991--2008 and over the longer interval 1991--2015 coincide,
within the error limits. A color excess is clearly seen in the
mean $K-L$ and $L-M$ colors, because $(K-L)< 0.^{m}2$ and $(L-M)<
0.^{m}0$ for normal (F5--K5) supergiants (Koornneef 1983).

Based on the data from Table~3, we can estimate the mean flux from
the dust shell at 5 $\mu$m. Below we assume that $K_{*}\approx
M_{*}$ for a late-type supergiants (Koornneef 1983). In 1991--2015
at $M_{*} \approx K_{*} \sim 3.^{m}6$ and the observed $M$
magnitude of $\sim 3.^{m}1$  (Table~3) the flux from the dust
shell was then

$F_{d}(M)\approx F_{obs}(M)-F_{*}(M)\approx (12.2-7.7)\cdot
10^{-17}\approx 4.5\cdot 10^{-17}$ (W cm$^{-2}$ $\mu$m$^{-1}$).

At the 2011 minimum (Table~3) the mean flux from the star at 5
$\mu$m was $F_{*}(M)\sim 6.14\cdot10^{-17}$ (W cm$^{-2}$
$\mu$m$^{-1}$) at $M_{*} \approx K_{*} \sim 3.^{m}85$, while the
total flux was $F_{obs}(M)\sim 10.5\cdot 10^{-17}$ (W cm$^{-2}$
$\mu$m$^{-1}$) for the observed magnitude $M \sim 3.^{m}27$.
Consequently, the flux from the dust shell was $F_{d}(M)\sim
4.4\cdot 10^{-17}$ (W cm$^{-2}$ $\mu$m$^{-1}$). Thus, the flux
from the dust shell at the 2011 minimum did not change, within the
error limits, and the observed reddening at $\lambda >2.2$ $\mu$m
is due to a decline in the star's brightness and an increase in
the relative contribution of the dust shell to the total flux.
Note also that in 1991--2015 the fraction of the stellar radiation
in the observed mean flux at 5 $\mu$m did not exceed 65\%.

The mean IR magnitudes and colors in 1991–-2008 and 2009–-2015
(outside the 2011 minimum) coincide, within the error limits, with
those obtained by averaging the data over the entire period of our
observations 1991–--2015 (Table~3). The mean fluxes from the dust
shell at 5 $\mu$m in all these time intervals and at the 2011
minimum are also almost identical, and, consequently, the mean
parameters of the dust shell can be assumed to have remain
unchanged over 25 years.

Bogdanov and Taranova (2009) performed a Fourier analysis of our
series of IR photometry in 1991--2008 and detected a period
$P=(237\pm1)^{d}$. In this paper we computed a periodogram (Fig.
8) for the complete series of observations of V1027 Cyg in
1991-–2015 in $J$ in the range of periods from 10$^{d}$ to
1000$^{d}$ using the computer code by Lenz and Breger (2005). The
observations at the 2011 minimum were excluded from consideration.
Analysis of the periodogram confirmed the period
$P=(237\pm1^{d})$. The two peaks of smaller height corresponding
to the 143$^{d}$ and 660$^{d}$ periods are a superposition of the
237$^{d}$ period with one year.

Figure 9 presents the phase $J$, $K$, and $M$ light and $J-H$,
$K-L$ and $L-M$ color curves constructed with the 237$^{d}$ period
and the epoch of minimum JD$_{min}$=24~48583.7. The observations
at the 2011 minimum drop out of the phase dependence.


\begin{table}

\caption{$JHKLM$ photometry for V1027 Cyg over the period
2009--2015}

\begin{tabular}{|c|c|c|c|c|c|}

\hline

JD&$J$&$H$&$K$&$L$&$M$\\

\hline

2454993.5&4.90&4.22&3.90&3.40&3.21\\
2455016.5&4.86&4.20&3.89&3.38&3.30\\
2455054.4&4.89&4.20&3.90&3.37&3.25\\
2455115.3&4.83&4.16&3.88&3.41&3.18\\
2455408.4&4.81&4.14&3.82&3.34&--\\
2455433.4&4.80&4.16&3.81&3.34&--\\
2455494.3&4.84&4.22&3.87&3.40&--\\
2455697.5&5.03&4.37&4.03&3.51&3.23\\
2455732.5&5.11&4.46&4.13&3.61&3.30\\
2455755.4&5.10&4.48&4.14&3.62&3.31\\
2455781.4&5.09&4.48&4.18&3.64&3.29\\
2455793.4&5.12&4.50&4.15&3.61&3.29\\
2455810.3&5.11&4.46&4.16&3.61&--\\
2456084.5&4.90&4.24&3.93&3.40&--\\
2456137.4&4.83&4.22&3.89&3.35&--\\
2456469.4&4.82&4.22&3.92&3.40&--\\
2456487.4&4.81&4.20&3.89&3.37&3.16\\
2456498.4&4.79&4.16&3.85&3.38&3.16\\
2456518.4&4.82&4.17&3.88&3.35&3.13\\
2456524.4&4.82&4.20&3.87&3.40&3.20\\
2456579.3&4.80&4.14&3.83&3.31&3.1\\
2456650.1&4.93&4.28&3.97&3.44&3.19\\
2456876.4&4.95&4.32&4.04&3.52&3.27\\
2456915.3&4.94&4.32&4.03&3.50&3.28\\
2456942.3&4.94&4.29&3.99&3.46&3.19\\
2456969.2&4.86&4.21&3.91&3.38&3.23\\
2457232.4&4.87&4.23&3.87&3.33&3.25\\
2457263.4&4.86&4.19&3.87&3.33&3.23\\
2457271.3&4.85&4.17&3.85&3.32&3.16\\
2457288.2&4.76&4.04&3.75&3.23&2.92\\
2457291.3&4.85&4.17&3.85&3.34&3.13\\
2457324.2&4.80&4.18&3.86&3.34&3.12\\

\hline
\end{tabular}

\end{table}

\begin{figure}

\includegraphics[scale=1.0]{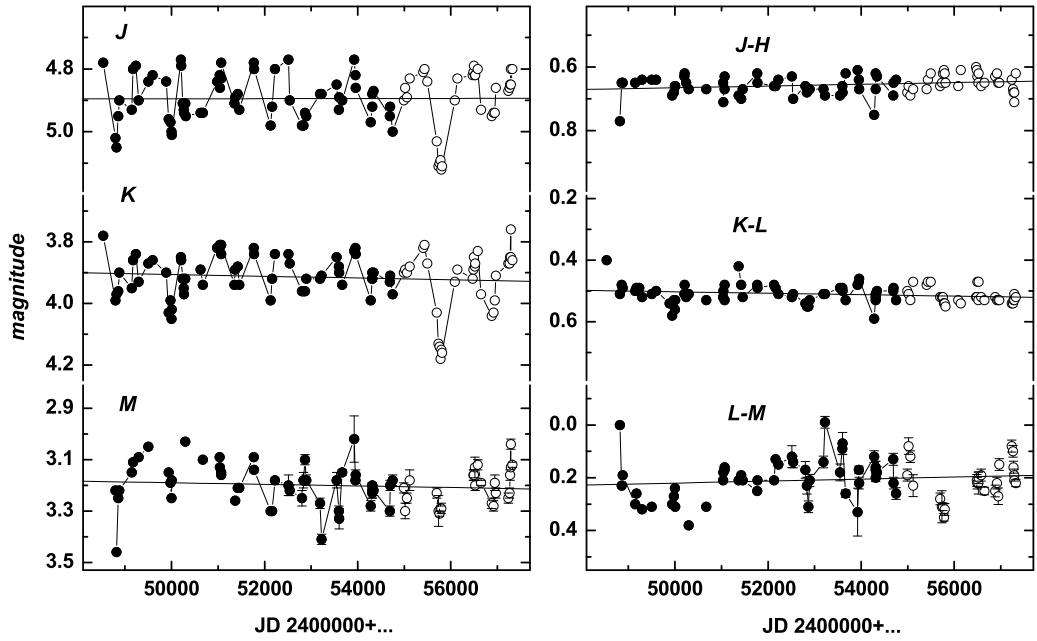}
\caption{$JKM$ light and $J-H$, $K-L$, and $L-M$ color curves of
V1027~Cyg from our Crimean observations from 1991 to 2015.}
\label{fig7}

\end{figure}


\begin{table}

\caption{Mean IR magnitudes and colors of V1027 Cyg in 1991--2008,
1991--2015, 2009--2015 (outside and at the 2011 minimum). The
extinction was taken into account with $E(B-V)=1.^{m}07$}

\begin{tabular}{c|c|c|c|c|c|c|c|c|c|c|c|c}

\hline Band &\multicolumn{3}{|c} {1991-2008}&\multicolumn{3}{|c}
{1991-2015}&\multicolumn{6}{|c} {2009-2015}\\

&\multicolumn{3}{|c}{}&\multicolumn{3}{|c}{}&\multicolumn{3}{|c}{outside
min 2011} &\multicolumn{3}{|c}{in min 2011}\\
\hline  & $\bar{m}$ & $\sigma_{\bar{m}}$ & N & $\bar{m}$ &
$\sigma_{\bar{m}}$ & N &
$\bar{m}$ & $\sigma_{\bar{m}}$ & N & $\bar{m}$ & $\sigma_{\bar{m}}$ & N\\
\hline
$J$&4.01&0.07&60&4.01&0.09&92&3.98&0.06&27&4.23&0.01&5\\
$H$&3.72&0.06&51&3.72&0.09&83&3.69&0.07&27&3.96&0.02&5\\
$K$&3.60&0.06&60&3.61&0.08&92&3.59&0.07&27&3.85&0.02&5\\
$L$&3.24&0.05&55&3.25&0.08&87&3.23&0.06&27&3.47&0.01&5\\
$M$&3.11&0.09&48&3.11&0.09&73&3.18&0.03&21&3.27&0.01&4\\
$J-H$&0.30&0.03&50&0.29&0.03&82&0.28&0.03&27&0.27&0.02&5\\
$H-K$&0.11&0.03&50&0.11&0.03&82&0.10&0.02&27&0.11&0.02&5\\
$K-L$&0.36&0.03&54&0.36&0.03&86&0.36&0.02&27&0.38&0.01&5\\
$L-M$&0.14&0.08&47&0.14&0.08&72&0.13&0.06&21&0.26&0.02&4\\

\hline
\end{tabular}

\end{table}


\begin{figure}

\includegraphics[scale=1.0]{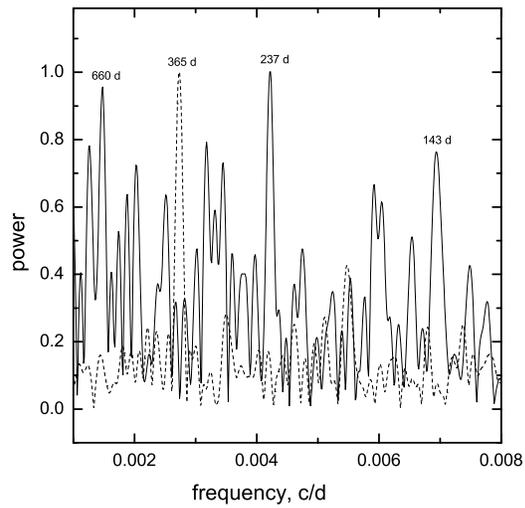}
\caption{Periodogram (solid line) and spectral window (dashed
line) from our $J$-band observations of V1027 Cyg from 1991 to
2015.} \label{fig8}

\end{figure}

\begin{figure}

\includegraphics[scale=1.0]{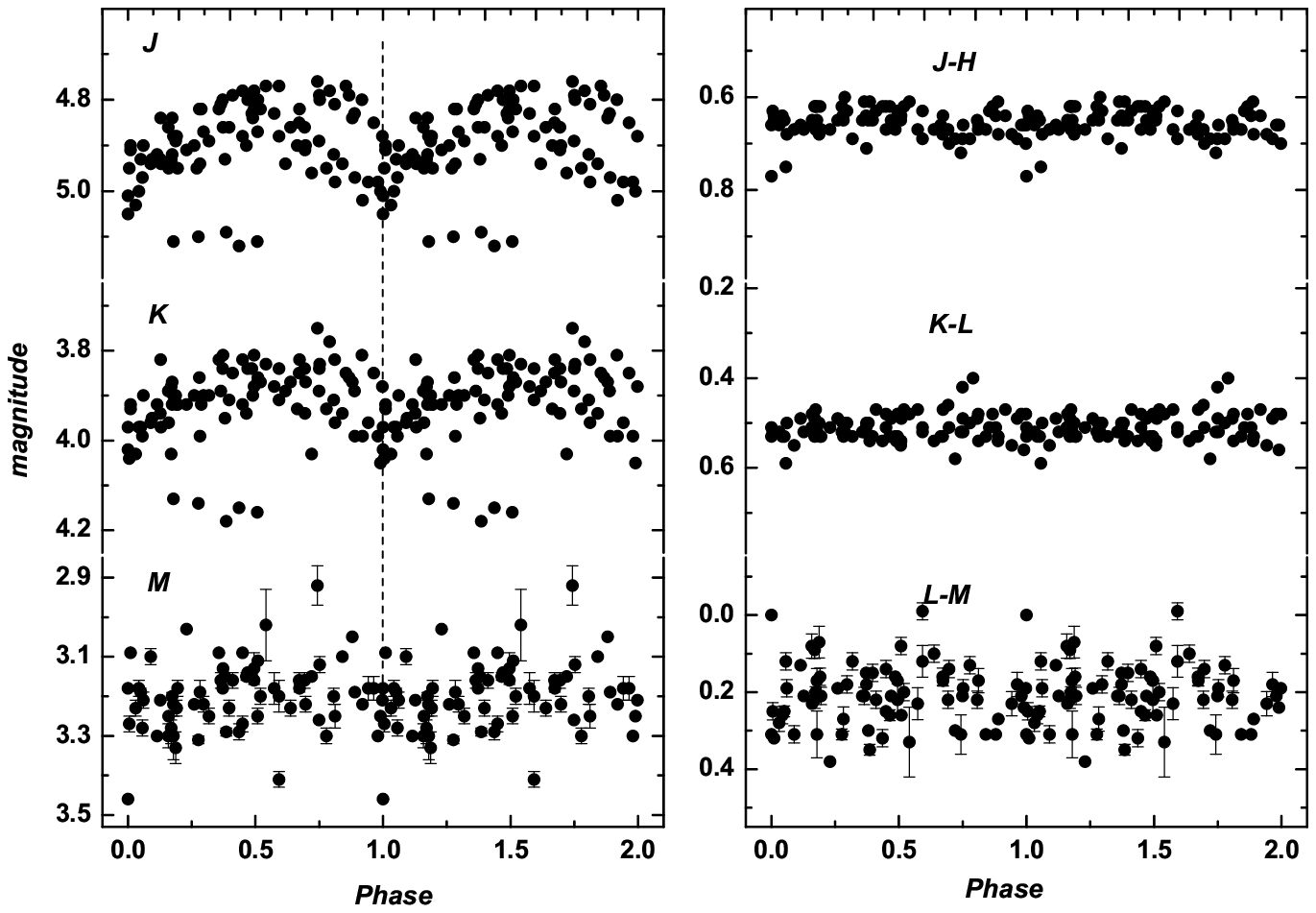}
\caption{$J$, $K$ and $M$ and $J-H$, $K-L$ and $L-M$ color
variations in V1027 Cyg with phase of the 237$^{d}$ period.}
\label{9}

\end{figure}

Figure 10 shows a color--magnitude $(J-H, J)$ diagram. It can be
seen from Fig. 10 that the bulk of the observed quantities lie
within the region bounded by the two straight lines A and B. Line
A reflects the possible $J$ brightness and $J-H$ color variations
in the star surrounded by a dust shell with a variable optical
depth whose particles are similar to interstellar ones. Line B is
the line of changes in the source's size at a constant optical
depth of the dust shell. The numbers near the crosses on line A
refer to $E'(B-V)=\Delta(J-H)/0.34$ and the initial values
(outside the dust shell) $J_{0}=4.^{m}67$ and
$(J-H)_{0}=0.^{m}62$.

The observed brightness and color variations in the near infrared
(1.25--1.65 $\mu$m), where the radiation completely belongs to the
stellar photosphere, can be explained by the combined effects from
(i) the variations in the optical depth of the circumstellar dust
shell, when $E'(B-V)$ can change from 0.$^{m}$0 to $\sim
0.^{m}35$, and (ii) the variations in the source's size. At the
observed mean $J$ brightness variations $\Delta J \sim 0.^{m}2$ at
a constant $J-H$ color the radial pulsations can reach 10\%.

\begin{figure}

\includegraphics[scale=1.0]{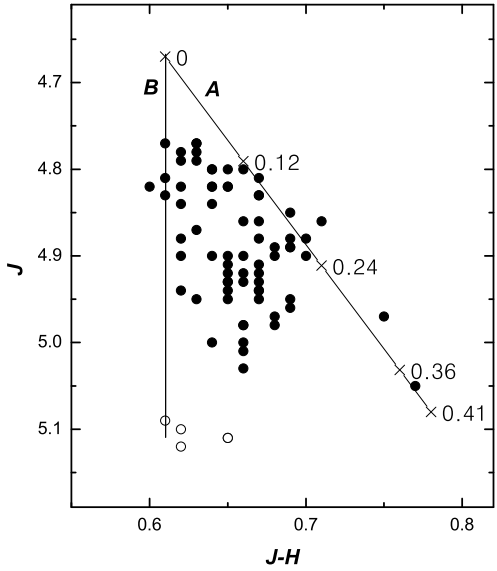}
\caption{($J-H,J$) diagram for the observed quantities of
V1027~Cyg. The bulk of the data are indicated by the circles. The
open circles refer to the 2011 minimum. Lines A and B indicate the
pattern of $J$ brightness and $J-H$ color variations in the
circumstellar dust shell and the $J$ brightness variations at a
given optical depth of the dust shell, respectively (see the
text). } \label{fig10}

\end{figure}

\subsection*{Spectral Energy Distribution for V1027 Cyg}

Our $UBVJHKLM$ photometry for V1027 Cyg allows us to construct the
spectral energy distribution for the star in the range 0.36--5.0
$\mu$m at various phases of the pulsation cycle. Consider two
epochs on the light curve of V1027 Cyg: JD$\sim$24 50672 is the
1997 minimum and JD$\sim$24 55440 is the 2010 maximum.

We dereddened the 1997 and 2010 photometry with $E(B-V)=1.^{m}07$,
constructed the spectral energy distribution for V1027 Cyg, and
fitted our data by the radiation of supergiants of different
spectral types (Fig. 11).

At maximum light $JD\sim$24~55440 the spectral energy distribution
for V1027 Cyg in a wide wavelength range, from 0.36~($U$) to
3.5~($L$) $\mu$m, is fitted quite satisfactorily by the radiation
of a G8I star for which the normal $U-B$ and $B-V$ colors were
taken from Strai\v{z}ys (1992), while the near-IR radiation is
presented in accordance with the data from Koornneef (1983). At
minimum light JD$\sim$24~50672 the spectral energy distribution
for V1027 Cyg corresponds to a K1 supergiant in the wavelength
range from 0.36 ($U$) to 2.2 ($K$) $\mu$m. The excess in $L$ and M
is associated with the contribution from the dust shell radiation.
Our estimates of the spectral types for V1027~Cyg are supported by
a comparison of the $UBVJHK$ data with the spectral energy
distribution for G8 and K1 supergiants from the library of stellar
spectra by Pickles (1998).

\begin{figure}

\includegraphics[scale=1.3]{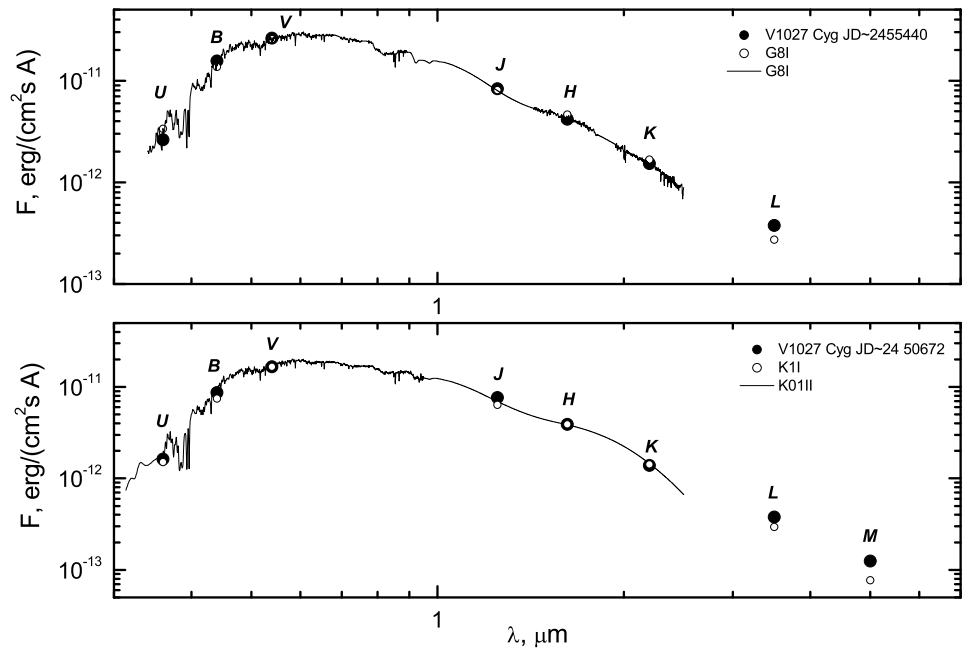}
\caption{Spectral energy distribution for V1027 Cyg corrected for
extinction with $E(B-V)=1.^{m}07$ and the data for normal
supergiants normalized to a wavelength of $\lambda$5500 \AA. The
upper and lower panels correspond to the maximum
(JD$\sim$24~55440) and the 1997 minimum (JD$\sim$24~50672),
respectively.}

\end{figure}

For V1027 Cyg at its deep 2011 minimum (JD$\sim$24~55810) we
failed to pick the spectral energy distribution of a normal
supergiant that would satisfy our photometric data in the entire
observed wavelength range. If the spectral type for the star is
taken to be K2--K3I, then a close correspondence of the dereddened
photometric data to the radiation of a normal K2--K3 supergiant is
observed in the range from 0.55 ($V$) to 3.5 ($L$) $\mu$m.
However, a significant excess of emission remains in the $U$ and
$B$ bands. When discussing our $UBV$ photometry, we have already
pointed out that the star became bluer at the 2011 minimum.

\section*{SPECTROSCOPIC OBSERVATIONS OF V1027 Cyg}

The spectroscopic observations of V1027 Cyg were carried out in
1995--2015 with the 1.25-m reflector at the Crimean Station of the
Sternberg Astronomical Institute, the Moscow State University,
using a spectrograph with a 600 lines/mm diffraction grating and a
long slit $\sim$ 4$^{\prime\prime}$ in width. The following CCD
array were used as the detector: ST-6 ($372\times274$-pixel CCD)
in 1995--2006 and ST-402 ($765\times510$-pixel CCD) in 2008--2015.
The formal spectral resolution was $\sim$ 5.5 \AA/pixel with the
ST-6 CCD and $\sim$ 2.2 \AA/pixel with the ST-402 CCD. A log of
spectroscopic observations is presented in Table~4, where the
dates of observations, the spectral range, and the $UBV$
photometry for the corresponding date are given.


\begin{table}
\caption{Log of spectroscopic observations and photometric data
for V1027 Cyg in 1995--2015}

\begin{tabular}{|c|c|c|c|c|c|}

\hline
Date &JD&Spectral range&$V$&$B-V$&$U-B$\\

\hline

23.08.1995&2449953&4400-7200&9.12&2.23&2.21\\
26.08.1995&2449956&4400-9200&9.15&2.23&2.24\\
02.09.1995&2449963&4400-9200&9.17&2.25&2.18\\
25.09.1995&2449986&4400-9200&9.23&2.21&2.42\\
15.08.1998&2451041&4400-7200&9.01&2.36&2.33\\
16.10.1999&2451468&4400-7200&9.27&2.28&2.16\\
01.09.2000&2451789&4400-9200&8.80&2.27&2.31\\
02.07.2003&2452823&4400-7200&9.04&2.28&2.27\\
02.08.2005&2453585&4400-7200&8.80&2.30&2.44\\
26.08.2006&2453974&4400-7200&8.68&2.21&2.28\\
11.07.2007&2454293&4400-7200&9.07&2.25&2.22\\
31.07.2008&2454679&4400-7200&8.87&2.23&2.16\\
19.08.2009&2455063&4400-9200&8.88&2.31&2.43\\
09.08.2010&2455418&4400-9200&8.70&2.20&2.21\\
21.10.2011&2455856&5000-9200&9.57&2.21&2.47\\
09.08.2015&2457244&4400-9200&9.10&2.48&2.64\\

\hline
\end{tabular}

\end{table}


The spectra were reduced with the standard CCDOPS code and the SPE
code (Sergeev and Heisberger 1993). We calibrated the fluxes based
on the spectra of the stars 18 Vul and 57 Cyg; their absolute
spectral energy distributions in the range $\lambda$4000-7650 \AA\
were taken from the spectrophotometric catalog by Voloshina et al.
(1982) and extended to $\lambda$9200 \AA\ using data from the
atlas of standard stellar spectra by Pickles (1985).

Figure 12 shows the times of our spectroscopic observations. We
took the spectra in 1995, 1998, 1999, 2003, 2005, 2006, 2007,
2011, and 2015 near the brightness minima, in 2006 at maximum
light, in 2008 in the middle of the descending branch of a deep
minimum, and in 2009 and 2010 in the middle of the ascending
branch of shallow minima.

All of the spectra, except those taken in 1995 and 2011, differed
insignificantly at different brightness levels, irrespective of
the pulsation phase. Figure 13 shows the absolutized and
extinction-corrected (with $E(B-V)=1.^{m}07$) spectrum of
V1027~Cyg on August 9, 2010, together with the spectrum of a G8I
supergiant from the library of stellar spectra by Pickles (1998).
The lines typical of late G or early K supergiants, in particular,
the numerous Fe I, Ti I, and Ca I absorptions and the Mg~I
$\lambda$5167-5183-5173 triplet, are represented in the wavelength
range 4500-8700 \AA. The main features of the spectrum for V1027
Cyg are the Ba II absorptions at $\lambda$4554, 5853, 6142 and
6497 and the IR Ca II triplet and O~I $\lambda$7771-7774 triplet
lines that are enhanced compared to normal supergiants. The
spectra exhibit strong Na~I D1 and D2 absorption and a DIB at
$\sim \lambda$6282.

TiO absorption bands appeared in the spectrum of V1027 Cyg near
the deep 1995 and 2011 minima, whose presence implies a spectral
type for the star later than K3. In the optical and near-IR ranges
these bands are prominent at $\lambda$5847, 6159, 6651, and 7065
\AA. In addition, in the 2011 spectrum an emission component is
clearly present in the H$\alpha$ profile. Figure 14 shows
fragments of the spectra for V1027 Cyg taken on September 25,
1995, and October 21, 2011, and, for comparison, the spectrum of
the post-AGB star IRAS 16476--1122 (sp = M1I) from the Appendix to
the spectral atlas by Su\'{a}rez et al. (2006). We estimated the
spectral type of V1027 Cyg in 1995 and 2011 from the intensity of
the TiO absorption bands to be M0--M1.


\begin{figure}

\includegraphics[scale=1.4]{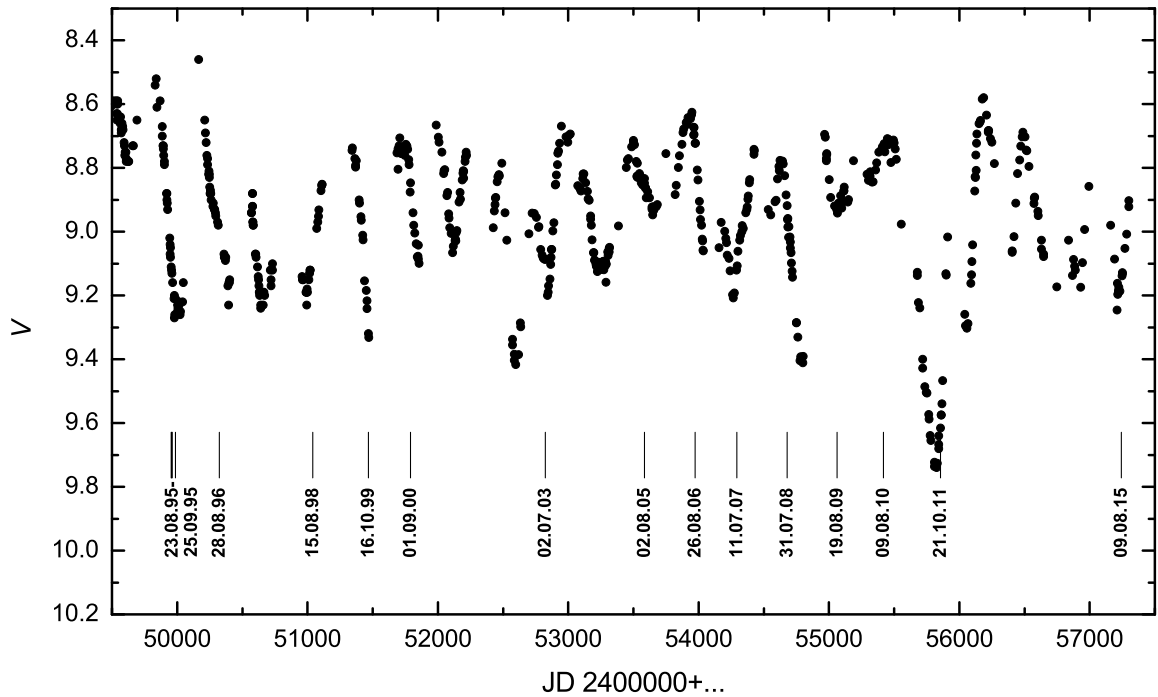}
\caption{$V$ light curve of V1027 Cyg from 1991 to 2015 with an
indication of the times of spectroscopic observations.}
\label{fig12}

\end{figure}



\begin{figure}

\includegraphics[scale=1.4]{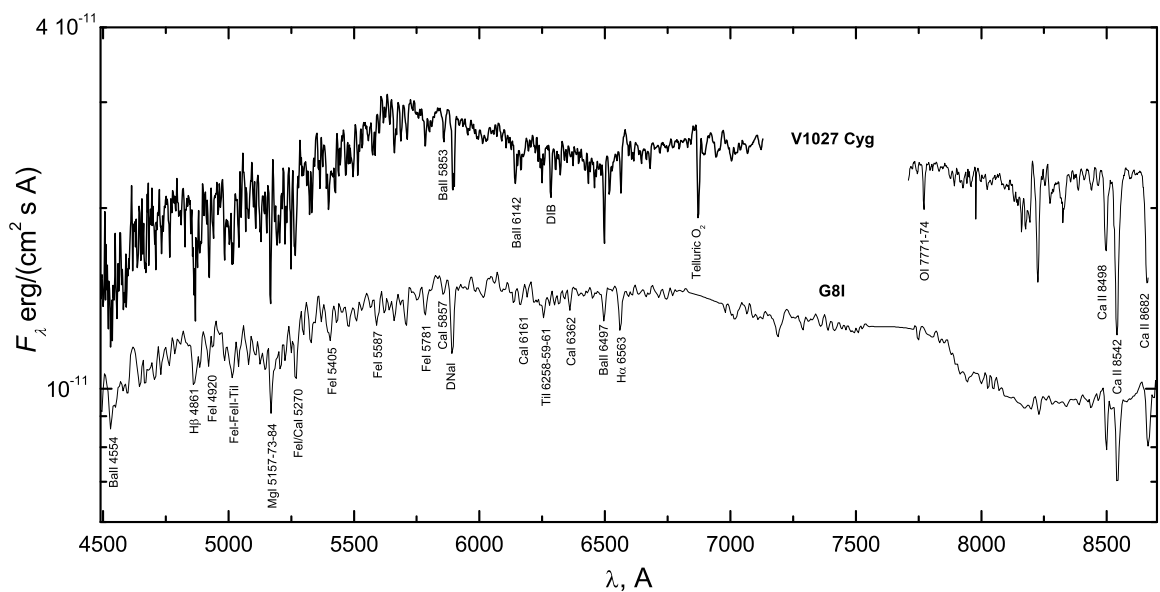}
\caption{Extinction-corrected spectrum of V1027 Cyg in absolute
energy units on August 9, 2010, and the spectrum of a G8
supergiant (Pickles 1998) whose level was shifted arbitrarily
along the vertical axis.} \label{fig13}

\end{figure}


It should be noted that the spectra with TiO molecular bands refer
to the 1995 and 2011 minima that differ by their photometric
characteristics. The 1995 minimum was "red"\ , while the 2011 one
was "blue"\ . In contrast, the 1995 and 1999 minima, which are
similar in their photometric behavior, a reddening as the
brightness declines, showed different spectra: one with TiO bands
and the other without them.


\begin{figure}

\includegraphics[scale=1.4]{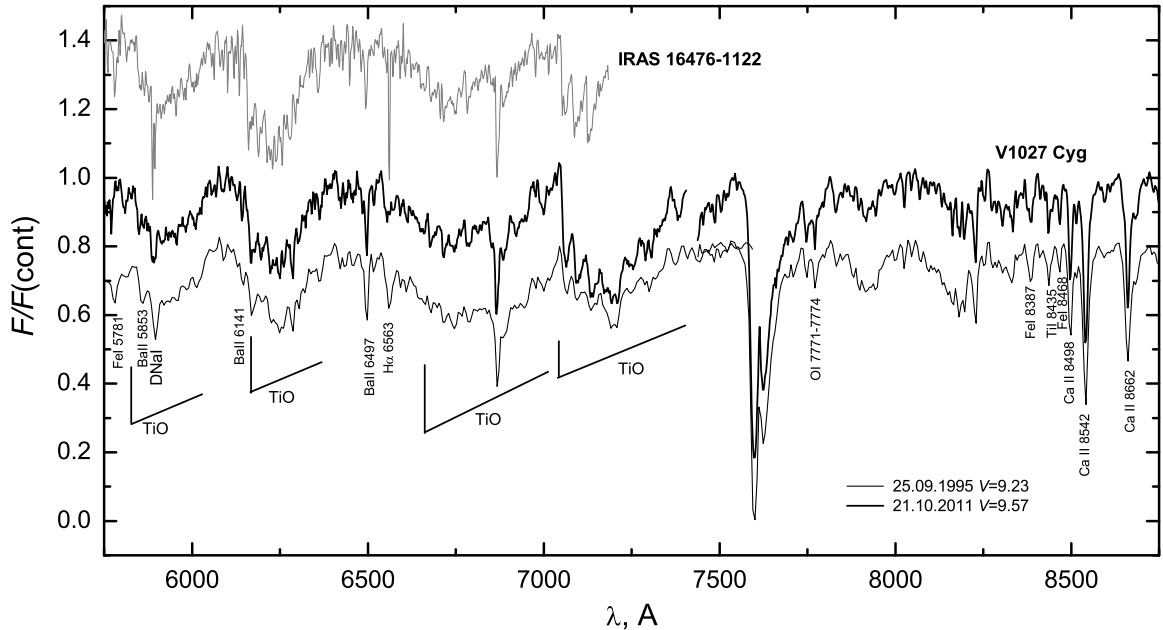}
\caption{Fragments of the spectra for V1027 Cyg (black lines) and
IRAS 16476--1122 (gray line). The spectra were normalized to the
continuum and arbitrarily shifted along the vertical axis.}
\label{fig14}

\end{figure}


We measured the equivalent widths of the strongest lines in the
spectrum of V1027 Cyg in the range $\lambda$5857-8670 \AA. The
equivalent widths did not show any correlations with both
brightness and color of the star, and their variations do not
exceed the measurement errors. Therefore, in Table~5 we give the
equivalent widths of the Na~D, Ba~II $\lambda$5857, $\lambda$6497,
H$\alpha$, O~I $\lambda$7771-7774 triplet, and Ca~II
$\lambda$8498, 8542, 8662 lines averaged over all spectra.


\begin{table}

\caption{Equivalent widths of absorption lines in the spectrum of
V1027 Cyg}

\begin{tabular}{|c|c|c|c|}
\hline

Line&$W_{\lambda}$, \AA &$\sigma W_{\lambda}$, \AA& $N$\\

\hline
Ba~II $\lambda$5857&   1.2&   0.3&   10\\
NaI $\lambda$5890-95&    3.3&   0.6&   11\\
Ba~II $\lambda$6497&   2.3&   0.6&   15\\
H$\alpha$ $\lambda$6563&     2.1&   0.3&   14\\
O~I $\lambda$7771-7774&   1.7&   0.2&    7\\
Ca~II $\lambda$8498&   3.9&   0.4&    8\\
Ca~II $\lambda$8542&   9.3&   0.9&    8\\
Ca~II $\lambda$8662&   7.2&   0.8&    8\\

\hline
\end{tabular}
\end{table}

In all spectrograms, except the 1999 one, the Ba~II $\lambda$6497
line is slightly stronger than H$\alpha$, in contrast to normal
supergiants of spectral types close to V1027 Cyg. The IR calcium
triplet absorptions are anomalously strong in the spectrum of
V1027 Cyg. The sum of the equivalent widths of all three lines
exceeds 20 \AA. The IR O~I triplet is not resolved in our spectra;
its equivalent width is also slightly overestimated compared to
normal supergiants.

\section*{DISCUSSION}

The $UBV$ observations of V1027 Cyg from 1991 to 2015 presented in
our paper revealed a photometric variability of the star on time
scales from 212 to 320 days with an amplitude up to 1$^{m}$ in
$V$, a bluing in some variations, and a reddening in others. The
complex photometric behavior of the star in the optical range may
suggest the existence of an additional factor that, apart from
pulsations, causes its photometric variability. The shock waves
associated with the radial pulsations contribute to an enhancement
of the mass loss by the star, which gives rise to an additional
dust component and to a change in the optical depth of the
circumstellar dust envelope. It can be assumed that the bluing of
the star is due to the scattering of stellar radiation by small
dust particles, while its reddening and fading are related not
only to a decrease in the temperature during pulsations but also
to extinction in the circumstellar dust envelope.

Our near-IR observations allowed the fraction of extinction in the
dust shell to be estimated, $E'(B-V)\leq 0.^{m}35$. Based on the
entire series of IR observations, except for the 2011 minimum, we
found a variability period of 237$^{d}$ that can be the pulsation
period.

The 2011 event deserves particular attention. It stands out by its
longer duration, the maximum variability amplitudes in all optical
and IR photometric minimum, when the $V$ brightness of the star
dropped by $\sim 1^{m}$, the $B-V$ and $U-B$ colors became bluer
(!) by a few hundredths of a magnitude compared to the maximum
brightness level. In $J$ at a variability amplitude of $0.^{m}25$
the $J-H$ color remained at the level corresponding to the star's
maximum light, and only in the range 2.2-5 $\mu$m did the colors
slightly increase.

The spectroscopy obtained at a time close to the photometric 2011
minimum revealed TiO molecular bands in the star's spectrum whose
intensity corresponded to spectral type M0--M1, while at maximum
light the star is classified as G7--G8I. The color differences for
the spectral types M0I and G8I are $\Delta (U-B)=0.^{m}83$,
$\Delta (B-V)=0.^{m}47$, $\Delta (J-H)=0.^{m}35$. This means that
the star must have reddened considerably, which was not observed.

The discrepancy between the spectroscopic and photometric
characteristics of the star is most probably due to the existence
of a circumstellar gas--dust shell around V1027 Cyg. As Klochkova
(2014) pointed out, the optical spectra of post-AGB stars differ
from the spectra of classical massive supergiants by the presence
of molecular bands superimposed on the spectrum of an F--G
supergiant.

V1027 Cyg is one of the coolest investigated post-AGB stars and so
far the only known oxygen-rich object of a late spectral type for
which long-term photometry is available. At the same time, V1027
Cyg differs from the variable carbon-rich post-AGB stars of late
spectral types (for example, IRAS 20000+3239 (Sp = G8I), IRAS
Z02229+6208 (Sp = G8.K0Ia), IRAS 05113+1347 (Sp = G8Ia); Hrivnak
et al. 2010) by longer periods and larger variability amplitudes.
Hrivnak et al. (2015) assumed that the period– temperature
relation for oxygen-rich objects could be steeper than that for
carbon-rich ones.

Figure 15 shows the period--temperature relation for 12
carbon-rich objects from Hrivnak et al. (2010) and for four
oxygen-rich stars: V887 Her, V1648 Aql, HD~331319, and V1027 Cyg.
The pulsation periods for V887 Her, V1648 Aql, and HD~331319 were
taken from Arkhipova et al. (2010, 2006), while the effective
temperatures for V887 Her and V1648 Aql were estimated by
Klochkova (1995) and Pereira et al. (2004), respectively. Two
effective temperature determinations exist for HD~331319:
$7750\pm100$ K (Arellano Ferro et al. 2001) and $7200\pm100$
(Klochkova et al. 2002).

V1027 Cyg with its parameters $P=237\pm1^{d}$ and $T_{eff}=5000$ K
lies above the sequence of carbon-rich objects in Fig. 15 and
confirms the assumption of Hrivnak et al. (2015) about a steeper
relation for oxygen-rich objects. The long period of V1027 Cyg may
suggest a high luminosity of the star.


\begin{figure}

\includegraphics[scale=1.0]{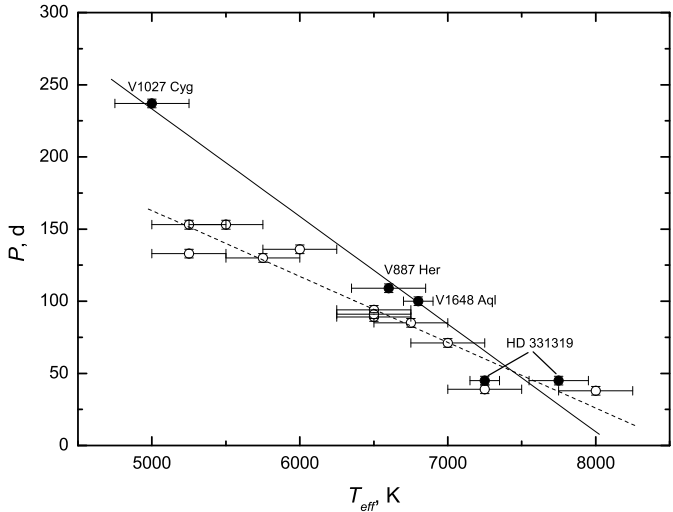}
\caption{Pulsation period versus effective temperature for 16
post-AGB stars. The open circles indicate 12 carbon-rich objects
from Hrivnak et al. (2010). The oxygen-rich objects are indicated
by the filled circles. The dashed line is a linear fit to the
period--temperature relation for the carbon-rich objects; the
solid line refers to the oxygen-rich objects.}

\end{figure}


Spectroscopic criteria also point to a high luminosity of V1027
Cyg. Klochkova et al. (2000) provided the absolute magnitude for
this star, $M_{V}\approx-7^{m}$, obtained from the calibration
relation between the IR O~I triplet equivalent width and stellar
luminosity (Arellano Ferro and Mendoza 1993).
$W_{\lambda}(\mathrm{O~I})=1.7\pm 0.2$ \AA\ from our paper also
corresponds to $M_{V}\approx-7^{m}$. However, it should be kept in
mind that the strengthening of the oxygen triplet can be caused
not only by the luminosity effect but also by an enhanced oxygen
abundance in an evolved star (Arellano Ferro et al. 2003).
Applying a different spectroscopic criterion, the dependence of
the Ba II $\lambda$ 5853 and 6141 \AA\ line equivalent widths on
stellar luminosity from Andrievsky (1998), also leads to a high
value of $M_{V}\approx -7^{m}\div -8^{m}$ (Klochkova et al. 2000).

Finally, the IR Ca~II triplet, whose intensity correlates with
$\log g$ (Jones et al. 1984), is considerably strengthened in the
spectrum of V1027 Cyg. The sum of the Ca II $\lambda$8498, 8542
and 8662 \AA\ line equivalent widths obtained from our
observations, $W_{\lambda}(\mathrm{Ca~T})=20.3\pm1.8$ \AA, exceeds
the maximum values for supergiants of similar chemical composition
and points to a high luminosity of the star (Cenarro et al. 2002).
For typical post-AGB stars of late spectral types (V354~Lac,
AI~CMi, IRAS~13313--5838) the IR Ca II triplet intensity is
approximately half that for V1027 Cyg (our unpublished data).

V1027 Cyg lies at a low Galactic latitude, $b=-0.^{\circ}36$. If
the distance to the star is taken to be $d\sim$1.26 kpc (Vickers
et al. 2015), then its height above the Galactic plane does not
exceed $z\sim$ 10 pc. This suggests that the star belongs to the
flat Galactic disk and, consequently, may be indicative of its
enhanced mass.

If V1027 Cyg is assumed to be a post-AGB star of enhanced mass,
then one would expect the observational manifestations of fast
evolution from it, in particular, a noticeable systematic change
in its mean visual brightness. Based on our photometric
observations over the last 25 years, we detected no trend in the
mean $UBV$ brightness. According to the first observations of the
star from Wachmann (1961), its photographic magnitude in the time
interval JD~24~32700-36900 (1948–-1959) changed within the range
$m_{pg}=10.^{m}5\div11.^{m}5$, while in 1991–-2015 the range of
$B$ magnitude variations was $B=10.^{m}6\div11.^{m}8$.
Consequently, the mean brightness of the star has not changed over
more than 60 years. The visual magnitude from the BD catalogue,
$9.^{m}1$, lies within the range of V brightness variations at the
present epoch.

There are estimates of the parameters $T_{eff}$ and $\log g$ for
V1027 Cyg in the literature. Klochkova et al. (2000) obtained two
sets of parameters by comparing the spectra with model
atmospheres: $T_{eff}$=5000 K, $\log g$=1.0 and $T_{eff}$=4900 K,
$\log g$=0.5. Meneses-Goytia et al. (2015) estimated
$T_{eff}$=5450 K and $\log g$=1.0 from their IR data. In both
cases, V1027 Cyg turns out to be near the track with the minimum
possible stellar mass of 0.53$M_{\odot}$ on the evolutionary
diagram  $\log T_{eff}-\log g$ from Miller Bertolami (2015) for
post-AGB stars.

Thus, our analysis of the new photometric and spectroscopic
observations together with the archival data does not yet give a
definite answer to the question about the status of V1027 Cyg
among other post-AGB stars. Some data suggest a high luminosity of
V1027 Cyg and, consequently, its enhanced mass, while other data
place the star in the region of low mass post-AGB objects.

\section*{CONCLUSIONS}

Our long-term observations of the semiregular pulsating variable
V1027~Cyg, a supergiant with an IR excess, led us to the following
conclusions about the photometric behavior and spectroscopic
peculiarities of this star.

In 1991–-2015 the star exhibited quasi-periodic brightness
variations with variable amplitudes and a changing light-curve
shape.

(1) We observed both small-amplitude variations with $\Delta
V$=0.$^{m}$2 in 1994, 2005, 2009 and variations with depths up to
0.$^{m}$6 as well as deep photometric minima with an amplitude up
to 1$^{m}$ in 1992 and 2011.

(2) The search for a period based on our $UBV$ photometry in
1991–-2015 revealed several periods from $P=212^{d}$ to
$P=320^{d}$ in different time intervals. The deepest 2011 minimum
does not satisfy any of these periods.

(3) The correlation of the $U-B$ and $B-V$ colors with the
brightness is not unambiguous. The fading of the star is
accompanied by its reddening in some variations and by its
noticeable bluing at the deepest 1992 and 2011 minima. There were
also brightness variations during which the reddest colors
occurred at minimum light, while subsequently $B-V$ and $U-B$
decreased and reached their lowest values in the middle of the
ascending branch, whereupon the star reddened as it brightened
further (the 2004 variation from JD$_{min}$=24~53250). In 2015 an
anomalous reddening of the star was observed as it faded.

(4) Our $JHKLM$ observations of V1027 Cyg in 1991–-2015 showed
quasi-periodic brightness variations with the most probable period
$P=237^{d}$ and amplitudes no greater than 0.$^{m}$2. The deep
2011 minimum that occurred in the optical range manifested itself
as a brightness decline by $0.^{m}25-0.^{m}30$ in $JHK$. This
minimum does not satisfy the period $P=237^{d}$.

(5) Based on our optical photometry, we determined the color
excess for V1027 Cyg: $E(B-V)\approx1.^{m}07$. Our IR observations
allowed us to estimate the fraction of circumstellar extinction
$E'(B-V)\leq 0.^{m}35$ under the assumption that the dust grains
in the circumstellar shell were similar to the interstellar ones.

(6) A joint analysis of our optical and IR photometry allowed us
to construct the spectral energy distribution for V1027 Cyg at
some phases of the pulsation cycle. The energy distribution was
shown to change and to correspond to spectral types from G8I to
K3I.

(7) Our low-resolution spectroscopy in the range
$\lambda$4400-9200 \AA\ showed the appearance of TiO emission
bands at the 1995 and 2011 minima. At the same time, no TiO bands
were observed near the deep 1992 minimum (Arkhipova et al. 1997)
and when the star faded considerably in 1998, 1999, and 2007.

(8) We measured the equivalent widths of the OI $\lambda$ 7774
\AA\ and IR calcium triple absorption lines, which are luminosity
criteria. The derived $W(\mathrm{O~I})=1.8\pm0.2$ \AA\ and
$W(\mathrm{Ca~T})=20.3\pm1.8$ \AA\ suggest a high luminosity of
the star.

Analysis of our long-term photometry for V1027 Cyg suggests that
the pulsational instability, along with the change in the optical
depth of the circumstellar dust envelope, is responsible for the
optical and near-IR variability of the star.

Simultaneous photometric observations and radial velocity
measurements for the star are needed to investigate the
variability due to its pulsations.

\section*{ACKNOWLEDGMENTS}

We used the SIMBAD database. This work was supported in part by
the Russian Foundation for Basic Research (project no.
06-02-16843).

\bigskip

REFERENCES
\bigskip

\begin{enumerate}

\item S. Andrievsky, Astron. Nachr. 319, 239 (1998).

\item A. Arellano Ferro and E. E. Mendoza, Astron. J. 106, 2516
(1993).

\item A. Arellano Ferro, S. Giridhar, and P. Mathias, Astron.
Astrophys. 368, 250 (2001).

\item A. Arellano Ferro, S. Giridhar, and E. Rojo Arellano, Rev.
Mex. Astron. Astrofi. s. 39, 3 (2003).

\item V. P. Arkhipova, V. F. Esipov, N. P. Ikonnikova, R. I.
Noskova, S. Yu. Shugarov, and N. A. Gorynya, Astron. Lett. 23, 690
(1997).

\item V. P. Arkhipova, N. P. Ikonnikova, R. I. Noskova, and S. Yu.
Shugarov, Astron. Tsirk., No. 1551 (1991).

\item V. P. Arkhipova, N. P. Ikonnikova, R. I. Noskova, and S. Yu.
Shugarov, Astron. Lett. 18, 418 (1992).

\item V. P. Arkhipova, N. P. Ikonnikova, G. V. Komissarova, and V.
F. Esipov, Astron. Lett. 32, 45 (2006).

\item V. P. Arkhipova, N. P. Ikonnikova, and G. V. Komissarova,
Astron. Lett. 36, 269 (2010).

\item M. B. Bogdanov and O. G. Taranova, Astron. Rep. 86, 850
(2009).

\item A. J. Cenarro, J. Gorgas, N. Cardiel, A. Vazdekis, and R. F.
Peletier, Mon. Not. R. Astron. Soc. 329, 863 (2002).

\item T. J. Deeming, Astrophys. Space Sci. 36, 137 (1975).

\item J. H. He, R. Szczerba, T. I. Hasegawa, and M. R. Schmidt,
Astrophys. J. Suppl. Ser. 210, 26 (2014).

\item E. Hog, C. Fabricius, V. V. Makarov, S. Urban, T. Corbin, G.
Wycoff, U. Bastian, P. Schwekendiek, and A.Wicenec, Astron.
Astrophys. 355, L27 (2000).

\item B. J. Hrivnak, Sun Kwok, and K. M. Volk, Astrophys. J. 346,
265 (1989).

\item B. J. Hrivnak, W. Lu, R. E. Maupin, and B. D. Spitzbart,
Astrophys. J. 709, 1042 (2010).

\item B. J. Hrivnak, W. Lu, and K. A. Nault, Astron. J. 149, 184
(2015).

\item H. L. Johnson, R. I. Mitchel, B. Iriarte, and W. Z.
Wisniewski, Comm. Lunar Planet. Lab. 4, 99 (1966).

\item J. E. Jones, D. M. Alloin, and B. J. T. Jones, Astrophys. J.
283, 457 (1984).

\item P. C. Keenan and R. McNeil, An Atlas of Spectra of the
Cooler Stars (Ohio State Univ. Press, Columbus, 1976).

\item V. G. Klochkova, Mon. Not. R. Astron. Soc. 272, 710 (1995).

\item V. G. Klochkova, T. V. Mishenina, and V. E. Panchuk, Astron.
Lett. 26, 398 (2000).

\item V. G. Klochkova, V. E. Panchuk, and N. S. Tavolzhanskaya,
Astron. Lett. 28, 49 (2002).

\item V. G. Klochkova, Astrophys. Bull. 69, 279 (2014).

\item J. Koornneef, Astron. Asrophys. 128, 84 (1983).

\item B. V. Kukarkin, P. N. Kholopov, Y. P. Pskovsky, et al.,
General Catalogue of Variable Stars, 3rd ed. (Nauka, Moscow, 1971)
[in Russian].

\item P. Lenz and M. Breger, Commun. Asteroseismol. 146, 53
(2005).

\item V. M. Lyutyj, Soobshch. GAISh No. 172, 30 (1971).

\item S. Meneses-Goytia, R. F. Peletier, S. C. Trager, J.
Falc\'{o}n-Barroso, M. Koleva, and A. Vazdekis, Astron. Astrophys.
582, A96 (2015).

\item M. M. Miller Bertolami, ASP Conf. Ser. 493, 133 (2015).

\item J. A. Orosz, J. R. Thorstensen and R. K. Honeycutt, Mon.
Not. R. Astron. Soc. 326, 1134 (2001).

\item C. B. Pereira, S. Lorenz-Martins, and M. Machado, Astron.
Astrophys. 422, 637 (2004).

\item A. J. Pickles, Astrophys. J. Suppl. Ser. 59, 33 (1985).

\item A. J. Pickles, Publ. Astron. Soc. Pasif. 110, 863 (1998).

\item J. T. Rayner, M. C. Cushing, and W. D. Vacca, Astrophys. J.
Suppl. Ser. 185, 289 (2009).

\item N. G. Roman, Spectral Classification and Multicolor
Photometry, Ed. by Ch. Ferenbach and B. Westerlund (D. Reidel,
Dordrecht, 1973), p. 36.

\item D. J. Schlegel, D. P. Finkbeiner, and M. Davis, Astrophys.
J. 500, 525 (1998).

\item S. G. Sergeev and F. Heisberger, A Users Manual for SPE
(Wien, 1993).

\item V. Strai\v{z}ys, Multicolor Stellar Photometry (Pachart,
Tucson, 1992).

\item V. Strai\v{z}ys, Metal-Deficient Stars (Mokslas, Vilnius,
1982) [in Russian].

\item O. Su\'{a}rez, P. Garc\'{i}a-Lario, A. Manchado, M.
Manteiga, A. Ulla, and S. R. Pottasch, Astron. Astrophys. 458, 173
(2006).

\item O. G. Taranova, V. I. Shenavrin, and A. M. Tatarnikov,
Astron. Lett. 355, 472 (2009).

\item B. Vandenbussche, D. Beintema, T. de Graauw, L. Decin, H.
Feuchtgruber, A. Heras, D. Kester, F. Lahuis, et al., Astron.
Astrophys. 390, 1033 (2002).

\item S. B. Vickers, D. J. Frew, Q. A. Parker, and I. S.
Boji\v{c}i\'{c}, Mon. Not. R. Astron. Soc. 447, 1673 (2015).

\item K. M. Volk and S. Kwok, Astrophys. J. 342, 345 (1989).

\item I. B. Voloshina et al., Spectrophotometry of Bright Stars
(Nauka, Moscow, 1982) [in Russian].

\item A. A. Wachmann, Berg. Abh. 6, 3 (1961).

\item S. Winfrey, C. Barnbaum, M. Morris, and A. Omont, Bull. Am.
Astron. Soc. 26, 1382 (1994).

\end{enumerate}

Translated by V. Astakhov

 \end{document}